\documentclass[11pt]{article}

\usepackage{hyperref}
\usepackage{amsmath,amssymb}
\usepackage{graphicx}
\usepackage{epstopdf}
\usepackage{bm}
\usepackage{color}

\newcommand{\be}{\begin{equation}}
\newcommand{\ee}{\end{equation}}
\newcommand{\bea}{\begin{eqnarray}}
\newcommand{\eea}{\end{eqnarray}}
\newcommand{\ba}{\begin{array}}
\newcommand{\ea}{\end{array}}

\newcommand{\tr}{\mbox{tr}}

\newcommand{\vap}{\varphi}

\newcommand{\bR}{{\bm{R}}}

\newcommand{\nn}{\nonumber}

\numberwithin{equation}{section}

\newcommand{\swedge}{\stackrel{\star}{\wedge}}

\begin{document}

\allowdisplaybreaks

\title{Charged rotating BTZ black holes in noncommutative spaces\\
and torsion gravity \bigskip}

\author{
        Shoichi Kawamoto$^*$,
        Koichi Nagasaki$^\star$
        and
	Wen-Yu Wen$^{*\dagger}$\bigskip\\
\footnotesize\tt kawamoto@cycu.edu.tw, 
nagasaki@cycu.edu.tw,
wenw@cycu.edu.tw
	\\
	$^*$\footnotesize\it 
Center for High Energy Physics and
Department of Physics,\\
\footnotesize\it 
 Chung-Yuan Christian University,
Chung-Li 320, Taiwan, R.O.C.\\
	$^\dagger$\footnotesize\it 
Leung Center for Cosmology and Particle Astrophysics,
National Taiwan University, Taipei 106, Taiwan
	}

\date{\today}

\maketitle

\bigskip

\begin{abstract}
\noindent\normalsize

We consider charged rotating BTZ black holes in noncommutative space by
use of Chern-Simons theory formulation of $2+1$ dimensional gravity.
The noncommutativity between the radial and the angular variables
is introduced through the Seiberg-Witten map for gauge fields,
and the deformed geometry to the first order in the noncommutative parameter
is derived.
It is found that the deformation also induces nontrivial torsion,
and 
the Einstein-Cartan theory appears to be a suitable framework to
investigate the equations of motion.
Though the deformation is indeed nontrivial,
the deformed and the original Einstein equations are 
found to be related by a rather simple coordinate transformation.

\end{abstract}


\clearpage

\setcounter{footnote}{0}
\section{Introduction}
\label{sec:introduction}

It is widely believed that at a very high energy scale, such as Planck or the string scale,
the notion of smooth geometry is no longer valid due to the effect of quantum gravity; namely quantum fluctuation of spacetime itself becomes significant and may not be treated as perturbation around a classical geometry.
Though we have not yet fully understand such a {\sl quantum geometry}, 
among those available proposals, the noncommutative geometry \cite{NC_geom} may capture some desired features of
it. In the quantum geometry, space-time coordinates are no longer regarded as $c$-numbers but
the ones obeying a specific quantum algebra, which naturally introduces a length scale serving as the UV cutoff.
Quantum field theory formulated on a noncommutative geometry also
exhibits various intriguing behaviors such as the UV/IR mixing
\cite{Minwalla:1999px} and stringy properties \cite{Yoneya:2000bt}.

The rich structure emerged in noncommutative quantum field theory has enticed several proposals to consider gravitational theory on
noncommutative space \cite{NC_gravity}.  Although it is not easy to
investigate concrete solutions due to their complicated structures, gravity in $2+1$ dimensions may be an exception.  For instance, the
Poisson brackets of $SL(2,\bR)$ were studied in Ref.\cite{Dolan:2006hv} and
families of deformation were found leading to a discrete spectrum for
time operator.  In Ref.\cite{Spallucci:2009zz}, an effective metric of
a noncommutative geometry was sourced by delocalized mass and charges due
to the minimal length.  In addition, $3D$ gravity is known
to have a description in terms of Chern-Simons theory \cite{CS_grav}.  
In this case, one may take advantage of the Seiberg-Witten map that relates a theory on
commutative space to a corresponding theory on a noncommutative space.  
To mention a few examples: the authors of \cite{Mukherjee:2006nd, Banerjee:2007th} used the Seiberg-Witten map to modify algebraic relation.  They found no first order correction as expected in the canonical treatment in the
noncommutative geometry, as long as classical torsion is excluded.
In Ref.\cite{Rivelles:2013ica}, the
ambiguity in the metric due to gauge transformation was discussed
and fixed by introducing nonminimal coupled scalars and a nontrivial
potential.  In Ref.\cite{Pinzul:2005ta}, it was argued that cosmological constant got quantized in the noncommutative Chern-Simons gravity.  

On the other hand, the three-dimensional anti-de Sitter (AdS) space admits the well-known black hole solution \cite{Banados:1992wn} and its charged counterpart \cite{Clement,Martinez:1999qi}.  In Ref.\cite{Banados:2001xw}, a constant gauge field was introduced in coupled with the Chern-Simon action and it amounted to mix mass and angular momentum in the original BTZ.  In Ref.\cite{ChangYoung:2008my}, a noncommutative
deformation in polar coordinates was introduced via the Seiberg-Witten
map and a noncommutative neutral BTZ black hole metric up to the first
order in  $\theta$ (noncommutative parameter) was obtained.  However,
this result appeared in conflict with that in Ref.\cite{Mukherjee:2006nd, Banerjee:2007th}
for its first order correction in metric.  Before we could solve the puzzle, 
it is useful to review their construction.


In Ref.\cite{ChangYoung:2008my}, a noncommutative deformation of a neutral rotating BTZ black
hole solution is investigated based on
a commutation relation in the polar coordinates, that is, 
$[r^2,\vap]=2i\theta$.\footnote{%
We will review the noncommutativity deformation in more detail in
Sec.\ref{sec:NC_gravity}.}
The solution is written in terms of Chern-Simons gauge fields and the noncommutative
deformation is introduced by the Seiberg-Witten map.
The resultant metric, to the first order in $\theta$, reads \footnote{%
We have corrected the sign mistakes in $f^2$
and $d\vap^2$ parts in Ref.\cite{ChangYoung:2008my}.}
\begin{align}
  ds^2 =&
-f^2 dt^2 + \hat{N}^{-2} dr^2
+2r^2 N^\phi dt d\varphi
+\bigg( r^2 - \frac{\theta B}{2} \bigg) d\varphi^2
+{\cal O}(\theta^2)
 \,,
\label{eq:neutral_def_BTZ}
\end{align}
with
\begin{align}
  N^\phi=& -\frac{r_+r_-}{\ell r^2} \,,\\
f^2 =& 
\frac{r^2 -r_+^2 -r_-^2}{\ell^2}  - \frac{\theta B}{2 \ell^2} \,,\\
\hat{N}^2 =&
\frac{1}{\ell^2 r^2}
\bigg[(r^2-r_+^2)(r^2-r_-^2) - \frac{\theta B}{2} \big( 2r^2-r_+^2-r_-^2\big) \bigg]
\,,
\end{align}
where $r_+$ represents the horizon radius of the undeformed metric
(the explicit forms are given in the Appendix \ref{sec:conventions}).
The noncommutative extension requires two extra $U(1)$ gauge fields
$B_\mu^{(\pm)}$, which are chosen as $B_\vap^{(\pm)}=B$ with a constant $B$.
Some properties of this deformed black hole solution are investigated 
in Ref.\cite{ChangYoung:2008my}, for instance, the locations of various types of horizon.

We, however, confirm that this metric satisfies the vacuum Einstein 
equation to the first order, 
$R_{\mu\nu}-\frac{1}{2}g_{\mu\nu}R+\Lambda g_{\mu\nu}=0+{\cal O}(\theta^2)$,
if the corrections of the metric in Ref.\cite{ChangYoung:2008my}
are taken into account.
This fact suggests that there should exist
another coordinate system in which the metric looks like a pure AdS${}_3$.
It turns out that, indeed, by making a coordinate transformation
\begin{align}
  r \rightarrow \tilde{r} +\frac{\theta B}{4\tilde{r}} \,,
\label{eq:coord_change}
\end{align}
and only keeping terms up to first order in $\theta$, the metric \eqref{eq:neutral_def_BTZ} comes back
to the undeformed BTZ black hole solution; namely the first order
correction can be eliminated.\footnote{%
There is a subtle issue about the regions covered by these
coordinates, which will be discussed in the Sec.\ref{sec:coordinate-change}.}
The angular part of the metric becomes $\tilde{r}^2 d\vap^2$, where $\tilde{r}$ is regarded as a standard radial coordinate.
Consequently, the deformed BTZ black
hole and the undeformed one are {\sl equivalent} up to the coordinate transformation \eqref{eq:coord_change}.
We remark that while the change is only made in $r$, the geometrical structure near the boundary would not be changed since $r$ and $\tilde{r}$ are asymptotically the same.  Thus, various mechanical and thermodynamic properties of black holes, such as the Hawking temperature, entropy, and orbital motion of particles, appear to be equivalent.
This equivalence may attribute to the fact that the vacuum solution of $2+1$
dimensional gravity with a negative cosmological constant is essentially
unique.
This motivates us to investigate a different class of solutions that are not
vacuum solutions.  In this paper, we shall explore the charged rotating BTZ
black hole solution in a noncommutative space.

The organization of the paper is as follows:
In the Sec.\ref{sec:NC_gravity}, a noncommutative deformation is
formulated by use of the Seiberg-Witten map in the Chern-Simons
framework of $2+1$ dimensional gravity.
We start with a charged rotating BTZ solution and obtain deformed
gauge fields, vielbeins, and spin connections.
In the Sec.\ref{sec:nonc-charg-rotat}, we investigate the properties of 
noncommutative charged rotating BTZ black hole solutions.
There appears nontrivial torsion and the deformed equations of motion 
are found to be nicely fitted in to the framework of
Einstein-Cartan theory of torsion gravity.
The relation between the deformed and the original solutions
through a coordinate change is expounded.
We conclude the paper with discussion and overview in
the Sec.\ref{sec:conclusion}.
The appendices are given to summarize our convention and to explain more technical details.

\section{Three dimensional gravity in noncommutative space}
\label{sec:NC_gravity}

\subsection{Noncommutativity in polar coordinates}
\label{sec:nonc-with-polar}

A noncommutative space is introduced by applying the following commutation
relations in the rectangular coordinates,
\begin{align}
  [x^\mu, x^\nu] = i \theta^{\mu\nu} \,.
\end{align}
Since timelike noncommutativity is known to have several difficulties,
such as acausality or nonunitarity \cite{timelikeNC},
we shall restrict our discussion to a purely spatial
noncommutativity; for example, $[x,y]=i\theta$ in $2+1$ dimensions, with $\theta$ being the parameter of
noncommutativity.
In this paper, however, the charged rotating BTZ black hole
in consideration is conveniently constructed in the polar coordinates
$(t,r,\vap)$ thanks to its azimuthal symmetry, one shall introduce a noncommutativity between
$r$ and $\vap$ coordinates instead.  As suggested in Ref.\cite{ChangYoung:2008my},  the noncommutative relation
\begin{align}
  [r^2, \vap]=2i\theta \,,
\label{eq:NC_rel1}
\end{align}
is a natural choice;
this is because the standard spatial noncommutative relation $[x,y]=i\theta$ can be recovered 
by use of the polar coordinates and \eqref{eq:NC_rel1} to the first
order in $\theta$, namely 
$[x,y]=[r \cos\vap, r \sin\vap]=i\theta + {\cal O}(\theta^2)$.
We thus adopt the noncommutative relation \eqref{eq:NC_rel1} and will consider a
$\theta$-deformed
charged rotating BTZ black hole solution.

\subsection{Charged rotating BTZ black hole solutions and Chern-Simons theory}
\label{sec:charged-rotating-btz}

\subsubsection{Chern-Simons formulation of Einstein-Maxwell theory}
\label{sec:chern-simons-form}

We start with Einstein-Maxwell theory in $(2+1)$ dimensions,
\begin{align}
  I=& I_\text{gravity} + I_\text{gauge}  \,,\\
I_\text{gravity} =& \frac{1}{16 \pi G} \int d^3 x \sqrt{-g} \big( R-2\Lambda  \big) \,,\qquad
I_\text{gauge} = -\frac{1}{4 \lambda^2} \int d^3 x \sqrt{-g} \mathfrak{f}_{\mu\nu} \mathfrak{f}^{\mu\nu}
 \,,
\label{eq:EM_theory1}
\end{align}
where $\lambda$ is the coupling constant of $U(1)$ gauge field $\mathfrak{a}_\mu$
, whose field strength is $\mathfrak{f}_{\mu\nu}= \partial_\mu \mathfrak{a}_\nu - \partial_\nu \mathfrak{a}_\mu$.
$I_\text{gravity}$ part of action can be rewritten by use of two $SU(1,1)\simeq SO(1,2)$ connection 1-forms (relevant conventions are summarized in  App.\ref{sec:conventions}),
\begin{align}
  A^{(\pm)a} = & \omega^a \pm \frac{1}{\ell} e^a 
\label{eq:gauge_vb_sc}
\end{align}
as the Chern-Simon terms
\begin{align}
  S=& I_{CS}[A^{(+)}] - I_{CS}[A^{(-)}] \,,\\
  I_{CS}[A] =& \frac{k}{4\pi} \int \tr \bigg[
A d A + \frac{2}{3} AAA \bigg]  \,,
\label{eq:CS_action}
\end{align}
where
the Chern-Simons level is given by $k= -\frac{\ell}{4G}$.

Using the definition of Hodge star,
$* (dx^\mu \wedge dx^\nu ) = \sqrt{|g|} \epsilon^{\mu\nu}{}_\rho dx^\rho$, 
the gauge part of action can be written as
\begin{align}
  I_\text{gauge}=& 
-\frac{1}{4 \lambda^2} \int d^3 x \sqrt{-g}  \mathfrak{f}_{\mu\nu} \mathfrak{f}_{\rho\sigma} g^{\mu\rho} g^{\nu\sigma}
=
-\frac{1}{2\lambda^2} \int \mathfrak{f} \wedge *\mathfrak{f} \,,
\end{align}
The equations of motion with respect to the variation of $A_\mu^a$ are
\begin{align}
\frac{k}{4\pi} \epsilon^{\mu\rho\sigma}  \bigg[ \partial_\rho A^{(\pm)a}_\sigma 
- \frac{\epsilon^a{}_{bc}}{2} A^{(\pm)b}_\rho A^{(\pm)c}_\sigma \bigg] =
\frac{\ell}{2} \cdot e {\cal T}^{\mu\rho} e_\rho^a \,,
\label{eq:eom_cBTZ}
\end{align}
where ${\cal T}^{\mu\rho}$ is the energy momentum tensor of $U(1)$
gauge field $\mathfrak{a}_\mu$, given by 
\begin{align}
{\cal T}_{\mu\nu}=&  \frac{1}{\lambda^2} \bigg[ \mathfrak{f}_{\mu\rho}
\mathfrak{f}_{\nu\sigma} g^{\rho\sigma} -\frac{1}{4}g_{\mu\nu} \mathfrak{f}^2 \bigg] \,.
\end{align}
In terms of the vielbeins and the spin connections,
the equations of motion can be also represented as
\begin{align}
&  \frac{k}{2\pi} \epsilon^{\mu\rho\sigma} \bigg[
\partial_\rho \omega^a_\sigma 
- \frac{\epsilon^a{}_{bc}}{2} \omega^b_\rho \omega^c_\sigma
- \frac{\epsilon^a{}_{bc}}{2 \ell^2} e^b_\rho e^c_\sigma \bigg]
=\ell \cdot e {\cal T}^{\mu\rho} e_\rho^a \,,
\label{eq:CS_eom1}
\\
&
\epsilon^{\mu\rho\sigma} \big( \partial_\rho e^a_\sigma - \epsilon^a{}_{bc} e^b_\rho \omega^c_\sigma \big) =0 \,.
\label{eq:torsion_free_comp}
\end{align}
The second equation is nothing but the torsion free condition, $T^a=De^a=de^a+\omega^a{}_b e^b  =0$,
while it is straight forward to see that the first one is the Einstein equation,
\begin{align}
  R_{\mu\nu} - \frac{1}{2}g_{\mu\nu} R - \frac{1}{\ell^2} g_{\mu\nu}
  =& 8\pi G {\cal T}_{\mu\nu} \,.
\label{eq:Einstein_eq}
\end{align}

\subsubsection{Charged rotating BTZ black hole}
\label{sec:charged-rotating-btz-1}

The charged rotating BTZ black hole solution is given in Ref.\cite{Martinez:1999qi}
as
\begin{align}
  ds^2 =& -f(r) dt^2 + \frac{dr^2}{f(r)} + r^2 \bigg( d\varphi - \frac{4G J}{r^2} dt \bigg)^2
\\=&
-h(r) dt^2 + \frac{dr^2}{f(r)} + r^2 d\vap^2
-\frac{2\gamma}{\ell} dtd\vap
 \,,
\label{eq:crBTZ_sol}
\\
f(r)=&
  -8G M +\frac{r^2}{\ell^2} + \frac{16 G^2 J^2}{r^2} - 8 \pi {G} Q^2 \ln r
=
\frac{1}{\ell^2} \bigg(-\alpha + r^2 + \frac{\gamma^2}{r^2} - \beta \ln r \bigg)
 \,,
\label{eq:f_CRBTZ}
\\
\mathfrak{f}_{tr} =& \frac{\lambda Q}{r} \,,
\label{eq:U1_flux}
\end{align}
where $Q$ is the electric charge of black hole
and we have defined
\begin{align}
\alpha= 8G M \ell^2 \,,
\quad
\beta= 8\pi G \ell^2 Q^2 \,,
\quad
\gamma= 4 G J \ell \,,
\end{align}
and $h(r)=f(r)-\frac{\gamma^2}{\ell^2 r^2}$ for our convenience.

From this metric, we shall choose a set of convenient, but not unique, vielbeins and spin connections as follows,
\begin{align}
  e^0 =& 
 \sqrt{h(r)} dt + \frac{\gamma}{\ell \sqrt{h(r)}} d\varphi
\,,\qquad
e^1 = \frac{1}{\sqrt{f(r)}}dr \,,\qquad
e^2=
 r \sqrt{\frac{f(r)}{h(r)}} d\varphi
\,,\\
  \omega^0=& -\frac{\gamma h'(r)}{2\ell r \sqrt{h(r)}} dt - \sqrt{h(r)} d\vap \,,
\qquad
\omega^1= \frac{\gamma h'(r)}{2\ell r h(r) \sqrt{f(r)}} dr \,,
\qquad
\omega^2= -\frac{h'(r)}{2} \sqrt{\frac{f(r)}{h(r)}} dt \,,
\end{align}
where the prime ${}'$ denotes the derivative with respect to $r$.
With this choice of vielbeins, the Chern-Simons gauge fields are written as (\ref{eq:gauge_vb_sc}):
\begin{align}
  A^{(\pm)0}=& \pm \frac{1}{\ell} \bigg( \sqrt{h(r)} \mp \frac{h'(r)}{2r} \frac{\gamma}{\sqrt{h(r)}} \bigg) dt
- \bigg( \sqrt{h(r)} \mp \frac{1}{\ell^2}
               \frac{\gamma}{\sqrt{h(r)}}\bigg) d\vap 
\label{eq:CS_g1}\,,\\
A^{(\pm)1}=&
\frac{1}{\ell \sqrt{f(r)}} \bigg( \frac{\gamma h'(r)}{2 r h(r)} \pm 1 \bigg) dr \,,
\label{eq:CS_g2}
\\
A^{(\pm)2}=& 
\frac{r}{\ell} \sqrt{\frac{f(r)}{h(r)}} 
\bigg( -\frac{\ell h'(r)}{2r} dt \pm d\vap \bigg) \,.
\label{eq:CS_g3}
\end{align}

\subsection{Noncommutative Chern-Simons theory}
\label{sec:nonc-chern-simons}

The Chern-Simons formulation of noncommutative three-dimensional
gravity has been considered in Ref.\cite{Banados:2001xw,Cacciatori:2002gq}.
In the Lorentzian version \cite{Cacciatori:2002gq}, it has been shown that
the theory involves two extra $U(1)$ gauge
fields $B^{(\pm)}_\mu$ and the gauge group becomes $U(1,1)\times U(1,1)$ rather than
$SO(1,2)\times SO(1,2)$ as in the commutative case, in which extra $U(1)$ fields will be decoupled.  The action of noncommutative Chern-Simons theory now reads
\begin{align}
  \hat{I}_{CS}[{\cal A}^{(\pm)}]=
\frac{k}{4\pi} \int \tr \bigg[
{\cal A}^{(\pm)} \swedge d  {\cal A}^{(\pm)}
+ \frac{2}{3} {\cal A}^{(\pm)} \swedge {\cal A}^{(\pm)} \swedge {\cal A}^{(\pm)}
\bigg] \,,
\end{align}
where
\begin{align}
  f \swedge g = \frac{1}{i!j!} f_{\mu_1 \cdots \mu_i} \star g_{\nu_1 \cdots \nu_j}
\, (dx^{\mu_1} \cdots dx^{\mu_i}) \wedge  (dx^{\nu_1} \cdots
  dx^{\nu_j}) \,,
\end{align}
and $\star$ represents the Moyal product
$f(x) \star g(x) =
e^{\frac{i}{2}\theta^{\mu\nu} \partial_\mu^x \partial_\nu^y} f(x) g(y)
\bigg|_{y \rightarrow x}$
 with an antisymmetric tensor $\theta^{\mu\nu}$.
The $SU(1,1)$ gauge fields $\hat{A}^{(\pm) a}$ ($a=0,1,2$) in 
the commutative Chern-Simons theory
, together with two extra $U(1)$ gauge fields
$\hat{B}_\mu$
form the new $U(1,1)$ gauge fields
\begin{align}
  {\cal A}_\mu^{(\pm)A} \tau_A = \hat{A}^{(\pm)a}_\mu \tau_a +  \hat{B}^{(\pm)}_\mu \tau_3 \,,
\end{align}
where ${\cal A}^{(\pm)3}=\hat{B}^{(\pm)}_\mu$.  We summarize convention for the
generators in App.\ref{sec:conventions}.

The equations of motion derived from the action $\hat{I}_{CS}$ read
\begin{align}
  \frac{\delta I_{CS}}{\delta \hat{A}_\mu^{(\pm)a}}=&
\mp \frac{k}{4\pi} \epsilon^{\mu\rho\sigma}
\bigg[
\eta_{ab}\partial_\rho \hat{A}_\sigma^{(\pm)b}
-\frac{\epsilon_{abc}}{2}\hat{A}_\rho^{(\pm)b}  \star \hat{A}_\sigma^{(\pm)c}
+\frac{i}{6}\eta_{ab}\big( \hat{A}_\rho^{(\pm)b} \star \hat{B}_\sigma^{(\pm)} +
                                               \hat{B}_\rho^{(\pm)} \star \hat{A}_\sigma^{(\pm)b} \big)
 \bigg]
\nn\\=& 0 \,,\\
  \frac{\delta I_{CS}}{\delta \hat{B}_\mu^{(\pm)}}=&
\pm \frac{k}{4\pi} \epsilon^{\mu\rho\sigma} 
\bigg[
\partial_\rho \hat{B}_\sigma^{(\pm)}
-\frac{i}{6} \eta_{ab} \hat{A}_\rho^{(\pm)a} \star \hat{A}_\sigma^{(\pm)b}
+ \frac{i}{2}\hat{B}_\rho^{(\pm)} \star \hat{B}_\sigma^{(\pm)}
\bigg]=0 \,.
\end{align}
In the commutative limit $\theta\rightarrow 0$, 
these equations boil down to the following decoupled equations of
motion,
\begin{align}
  F^{(\pm)a}=0 \,, \qquad
dB^{(\pm)}=0 \,.
\end{align}

\subsection{Coupling of matter fields to noncommutative gravity}
\label{sec:coupl-matt-fields}

Here we briefly discuss the coupling of the Abelian gauge field $\mathfrak{a}_\mu$
to the noncommutative gauge field ${\cal A}_\mu^{(\pm)A}$.
The straightforward extension of Maxwell action reads
\begin{align}
  \int d^3x \, \sqrt{-\hat{g}} \hat{\mathfrak{f}}_{\mu\nu} \hat{\mathfrak{f}}_{\rho\sigma}
\hat{g}^{\mu\rho} \hat{g}^{\nu\sigma} \big|_{\star} \,,
\end{align}
where the products between fields are understood to be the star product,
and the field strength is defined as
\begin{align}
  \hat{\mathfrak{f}}_{\mu\nu} =& 
\partial_\mu \hat{\mathfrak{a}}_\nu - \partial_\nu \hat{\mathfrak{a}}_\mu
+\hat{\mathfrak{a}}_\mu \star \hat{\mathfrak{a}}_\nu \,,
\end{align}
with $\hat{\mathfrak{a}}_\mu$ being a noncommutative extension of $U(1)$ gauge field.
The metric $\hat{g}_{\mu\nu}= \eta_{ab} \hat{e}_\mu{}^a \star \hat{e}_\nu{}^b$ is
given by the noncommutative extension of vielbeins,
 $\hat{e}_\mu{}^a=\frac{\ell}{2}\big( \hat{A}_\mu^{(+)a} - \hat{A}_\mu^{(-)a} \big)$.
As a result, the Maxwell action is highly nonlinear in terms of $\hat{A}_\mu^{(\pm)a}$.

On top of this nonlinearlity, the standard Maxwell action in a curved background
poses a question: general coordinate transformation is given by a field dependent
gauge transformation and this action is in general not fully $SU(1,1)$ gauge invariant.
If we want to maintain the full gauge invariance, we need to write the matter part coupling
in terms of ${\cal A}^{(\pm)}$ and $\mathfrak{a}$ in such a way that 
the obtained action keeps gauge invariance intact and also
comes back 
to \eqref{eq:EM_theory1} in the commutative limit.
This is a fairly nontrivial problem without immediate answer.
We thus take the following strategy:
the noncommutative gravity is introduced via the Seiberg-Witten map in the formulation of Chern-Simons theory,
while the matter part is treated as being coupled to the noncommutative gravity through
the gravitational degrees of freedom, namely, 
$\hat{g}_{\mu\nu}$, $T_{\mu\nu}{}^\rho$ (torsion), and $\hat{B}^{(\pm)}$ instead of
${\cal A}^{(\pm)}$.\footnote{%
We just choose metric and torsion as the fundamental degrees 
of freedom in noncommutative gravity, though
it is also possible to use $\hat{e}_\mu{}^a$ and $\hat{\omega}_\mu{}^a{}_b$ instead.}

We may try to write down possible coupling terms, which are classified by the numbers of derivatives and torsion tensors.  
It is known that gauge field cannot minimally couple to torsion in a gauge-invariant way, therefore we shall consider the non-minimal coupling.
The possible terms of lower dimensions are (we omit the symbol of the star product)
\begin{align}
I'_\text{gauge}=&  \xi_0^{(\pm)} \int \hat{\mathfrak{a}} \wedge d\hat{B}^{(\pm)}
+ \xi_1^{(\pm)} \int d^3 x \, \sqrt{-\hat{g}} \hat{\mathfrak{f}}_{\mu\nu} (d\hat{B}^{(\pm)})^{\mu\nu}
+ \int d^3 x \, \sqrt{-\hat{g}} \hat{\mathfrak{f}}_{\mu\nu} K^{\mu\nu} \,,
\end{align}
where the first term is gauge invariant up to a surface term.
The tensor $K^{\mu\nu}$ is torsion dependent and reads,
\begin{align}
  K^{\mu\nu}=&
K^{\mu\nu}_1+K_2^{\mu\nu} \,,\\
K^{\mu\nu}_1=&
\tilde\zeta_1 T^{\rho\sigma\mu} T_{\rho\sigma}{}^\nu
+\tilde\zeta_{2} T^{\rho\mu\sigma} T_{\rho}{}^\nu{}_\sigma
+\tilde\zeta_{3} T^\mu T^\nu
+\tilde\zeta_{4} T_\epsilon^{\mu\rho} T_\epsilon^{\nu}{}_\rho
+\tilde\zeta_{5} T_{\epsilon'}^{\mu\rho} T_{\epsilon'}^{\nu}{}_\rho
\,,\\
K^{\mu\nu}_2=&
\zeta_1 \epsilon^{\mu\rho\sigma} T_{\rho\sigma}{}^\nu
+\zeta_2 T_{\epsilon}^{\mu\nu}
+\zeta_3 T_{\epsilon'}^{\mu\nu}
+\zeta_4 \partial_\rho T^{\mu\nu\rho}
+\zeta_5 \partial_\rho T^{\mu\rho\nu}
+\zeta_6 \partial^\mu T^\nu \nn\\&
+\zeta_7 T^{\mu\nu\rho} T_\rho
+\zeta_8 T^{\mu\rho\nu} T_\rho
+\zeta_{9} T^{\rho\mu\sigma} T_{\rho\sigma}{}^\nu
+\zeta_{10} \epsilon^{\mu\nu\rho} T_{\sigma\tau\rho} T_\epsilon^{\sigma\tau}
+\zeta_{11} \epsilon^{\mu\nu\rho} T_{\sigma\tau\rho} T_{\epsilon'}^{\sigma\tau}
\nn\\&
+\zeta_{12} \epsilon^{\mu\nu\rho} T_{\rho \sigma\tau} T_\epsilon^{\sigma\tau}
+\zeta_{13} \epsilon^{\mu\nu\rho} T_{\rho \sigma\tau} T_{\epsilon'}^{\sigma\tau}
+\zeta_{14} \epsilon^{\mu\nu\rho} T_{\rho \sigma\tau} T_\epsilon^{\tau\sigma}
+\zeta_{15} \epsilon^{\mu\nu\rho} T_{\rho \sigma\tau} T_{\epsilon'}^{\tau\sigma}
\nn\\&
+\zeta_{16} T_\epsilon^{\mu\rho} T_{\epsilon'}^{\nu}{}_\rho
+\zeta_{17} T_{\epsilon}^{\mu}{}_\rho T_{\epsilon}^{\rho\nu}
+\zeta_{18} T_{\epsilon'}^{\mu}{}_\rho T_{\epsilon}^{\rho\nu}
+\zeta_{19} T_{\epsilon}^{\mu}{}_\rho T_{\epsilon'}^{\rho\nu}
+\zeta_{20} T_{\epsilon'}^{\mu}{}_\rho T_{\epsilon'}^{\rho\nu}
\nn\\&
+\epsilon^{\rho_1 \rho_2 \rho_3} T_{\rho_1 \rho_2 \rho_3}
\big[
\zeta_1' \epsilon^{\mu\rho\sigma} T_{\rho\sigma}{}^\nu
+\zeta_2' T_{\epsilon}^{\mu\nu}
+\zeta_3' T_{\epsilon'}^{\mu\nu} \big] \,,
\end{align}
where $T_\mu=\delta^\rho_\sigma T_{\mu\rho}{}^\sigma$ is the torsion vector (the trace of torsion),
$T_\epsilon^{\mu\nu}=\epsilon^{\mu\rho\sigma}T_{\rho\sigma}{}^\nu$, and
$T_{\epsilon'}^{\mu\nu}=\epsilon^{\mu\rho\sigma}T^\nu{}_{\rho\sigma}$.
Note that for $T_\epsilon^{\mu\nu}$ and $T_{\epsilon'}^{\mu\nu}$ the order of indices is important.
Here, only the terms up to mass dimension 2 are presented.

In the commutative limit $\theta\rightarrow 0$, we expect that $I'_\text{gauge}=0$
or $B_\mu^{(\pm)}$ and torsion are decoupled from the usual Einstein-Maxwell part. 
Since $K_1^{\mu\nu}$ becomes symmetric in the commutative limit,
the coupling constants $\tilde{\zeta}_i$ ($i=1,\cdots,5$) can be nonvanishing.
On the other hand, the other terms, if not vanished as $\theta \to 0$, would have affected the leading order solution.  
We may take the coupling constants $\xi_0^{(\pm)}$ and $\zeta_i$ to be proportional to
$\theta$ as a simple choice.

In the following subsection,
we first introduce the noncommutativity to gravity part via Seiberg-Witten map.
The desirable deformation of the matter part will be discussed later in the Sec.\ref{sec:matt-energy-moment}.

\subsection{Seiberg-Witten map}
\label{sec:seiberg-witten-map-1}

The Seiberg-Witten map \cite{Seiberg:1999vs} is introduced as a map
between gauge theories on commutative and noncommutative geometries.
As shown in Ref.\cite{Grandi:2000av}, Chern-Simons
theory has a peculiar feature under the map; the form of the action remains
unchanged (up to surface terms), and we can simply replace the ordinary products
with the Moyal products.  
This property suggests that at least for the part of Chern-Simons
action, a solution for the equations of motion can be mapped into its noncommutative counterpart.

We now consider the Seiberg-Witten map  based on the
radius-angle commutation relation \cite{ChangYoung:2008my}
\begin{align}
  [ \hat{R} ,  \hat{\vap}] = 2i \theta 
\end{align}
where $\hat{R}= \hat{r}^2$.
Namely, $\theta^{R\vap}=-\theta^{\vap R}=2 \theta$ and the other
components are all zero.
The convention is fixed in App.\ref{sec:seiberg-witten-map}, and
the correction term from the Seiberg-Witten map is
\begin{align}
  A'_\mu(A)=&
-\frac{i}{4} (2\theta) \bigg[
\frac{1}{2} \eta_{ab} A_R^a (\partial_\vap A_\mu^b + F_{\vap \mu}^b) \bm{1}
-\frac{1}{2} \eta_{ab} A_\vap^a (\partial_R A_\mu^b + F_{R \mu}^b) \bm{1}
\nn\\& \hskip3em
+i (A_R^a \tau_a + B_R \tau_3) ( \partial_\vap B_\mu  + F_{\vap \mu}^{(B)} )
-i (A_\vap^a \tau_a + B_\vap \tau_3) ( \partial_R B_\mu  + F_{R \mu}^{(B)} )
\nn\\& \hskip3em
+i B_R (\partial_\vap A_\mu^b + F_{\vap \mu}^b)\tau_b
-i B_\vap (\partial_R A_\mu^b + F_{R \mu}^b)\tau_b
\bigg] \,.
\label{eq:SWmap2}
\end{align}

Since the noncommutative version of Chern-Simons theory has
two extra gauge fields $B^{(\pm)}_\mu$, we need to give their forms
in the commutative case, where they have vanishing field strength, that is, $dB^{(\pm)}=0$.
We consider the simplest case with $B^{(\pm)}_\mu=Bd\vap$ for a constant $B$.
Then the Seiberg-Witten map now reads 
\begin{align}
  A_\mu^{(\pm)a\prime}=&
-\frac{\theta B}{2} \big[ \partial_R A_\mu^{(\pm)a} + F_{R\mu}^{a} \big] \,,\\
B_\mu^{(\pm)\prime}=& -\frac{\theta}{2} \eta_{ab} \big[ A_R^{(\pm)a} F_{\vap\mu}^b
-A_\vap^{(\pm)a} F_{R\mu}^b - A_\vap^{(\pm)a} \partial_R A_\mu^{(\pm)b} \big] \,.
\end{align}
By applying this map to the gauge fields \eqref{eq:CS_g1}--\eqref{eq:CS_g3},
to the first order in $\theta$, the noncommutative gauge fields are
\begin{align}
  \hat{A}^{(\pm)0}_t
=& \pm \frac{1}{\ell} \bigg( \sqrt{h} \mp h' \frac{\gamma}{2r \sqrt{h}} \bigg)
-\theta B \frac{(2r^2 -\beta)^2 \gamma
\pm 2\ell^2 (2r^4 -\beta r^2 \mp 4\beta \gamma) h}{16 \ell^5 r^4 h^{3/2}} 
\,,
\label{eq:A0t}
\\
  \hat{A}^{(\pm)0}_{\vap}
=&
- \bigg( \sqrt{h} \mp \frac{1}{\ell^2} \frac{\gamma}{\sqrt{h}}\bigg) 
+ \theta B \frac{\pm \gamma (2r^2-\beta) +2\ell^2 (r^2-\beta)h}{8 \ell^4 r^2 h^{3/2}}
\label{eq:A0f}
\,,\\
\hat{A}^{(\pm)1}_r=&
\frac{1}{\ell \sqrt{f}} \bigg( \frac{\gamma {h}'}{2r h(r)} \pm 1 \bigg) 
+\frac{\theta B}{32 \ell^7 r^6 h^2 f^{3/2}}
 \bigg[ 4\gamma^3(2r^2-\beta)^2+2\ell^2 h \big[3\gamma
                     r^2(2r^2-\beta)^2-4\beta \gamma^3
\nn\\& \hskip8em 
+2\ell^2 r^2 h
                     \big( \pm r^2(2r^2-\beta)+\gamma(2r^2-3\beta) \pm
                     2\ell^2 r^2 h \big) \big] \bigg]
 \,,
\label{eq:A1r}
\\
\hat{A}^{(\pm)2}_t=& 
- \frac{h'}{2} \sqrt{\frac{f}{h}}
+\theta B \frac{-\gamma^2 (2r^2-\beta)^2 +2\ell^2 h \big[  4 \beta \gamma^2 + 2\ell^2 r^2h (r^2+\beta)\big]}{16\ell^6 r^5 h^{3/2} {f}^{1/2}}
\label{eq:A2t}
\,,\\
\hat{A}^{(\pm)2}_\vap=& 
\pm \frac{r}{\ell} \sqrt{\frac{f}{h}}
\mp \theta B \frac{2 \beta \gamma (\pm \ell^2 h + \gamma)+ 4r^2 (\ell^4 h^2 -\gamma^2)}{16\ell^5 r^3 h^{3/2} {f}^{1/2}}
\label{eq:A2f}
 \,,\\
  \hat{B}^{(\pm)}=& \bigg( B +\frac{\beta \theta}{4\ell^2 r^2} \bigg) d\vap
-\beta\theta \frac{\pm r^2 + 2\gamma}{4\ell^3 r^4} dt \,,
\label{eq:Btf}
\end{align}
where the prime ${}'$ denotes the $r$ derivative.

In the following section, we discuss black hole solutions in
noncommutative gravity based on these expressions.
Note that since gauge fields are all functions of $r$ only,
we can again replace $\star$ product with a usual product.

\section{Noncommutative charged rotating BTZ black holes and torsion
  gravity }
\label{sec:nonc-charg-rotat}

\subsection{$\theta$-deformed metric}
\label{sec:nonc-metr}

In Sec.\ref{sec:NC_gravity}, we have derived the noncommutative
Chern-Simons gauge fields \eqref{eq:A0t}--\eqref{eq:Btf}.
From them we can reconstruct noncommutative vielbeins and
spin connections as follows:
\begin{align}
  \hat{e}^a =& \frac{\ell}{2} (\hat{A}^{(+)a} -  \hat{A}^{(-)a}  ) \,,
\qquad
  \hat{\omega}^a = \frac{1}{2} (\hat{A}^{(+)a} +  \hat{A}^{(-)a}  ) \,,
\end{align}
and the explicit forms are
\begin{align}
  \hat{e}^0 =& \bigg( \sqrt{h} - \theta B \frac{2r^2-\beta}{8\ell^2 r^2 \sqrt{h}}\bigg) dt
+\frac{\gamma}{\ell \sqrt{ h}} \bigg(1  + \theta B \frac{2r^2-\beta}{8h \ell^2 r^2} \bigg) d\varphi 
\label{eq:e0_def}\,,\\
\hat{e}^1=&  \bigg[ \frac{1}{\sqrt{f}} 
+\theta B \frac{2\ell^2 h +2r^2-\beta}{8 \ell^2 r^2 {f}^{3/2}} \bigg] dr \,,\\
\hat{e}^2=&  \bigg[ r \sqrt{\frac{f}{h}}  
- \theta B  \frac{2\ell^4 r^2 h^2  - (2r^2-\beta)\gamma^2}{8 \ell^4 r^3 h^{3/2} {f}^{1/2}}
  \bigg] d\varphi \,,\\
  \hat{\omega}^0=&
\bigg[ 
-\frac{\gamma h'}{2\ell r \sqrt{h}}
+\theta B \gamma \frac{8\ell^2 \beta h - (2r^2-\beta)^2}{16 \ell^5 r^4 h^{3/2}}
\bigg]dt
+\bigg[
-\sqrt{h}+\theta B \frac{r^2-\beta}{4 \ell^2 r^2 \sqrt{h}}
\bigg] d\vap \,,\\
  \hat{\omega}^1=&
\bigg[
\frac{\gamma h'}{2\ell r h \sqrt{f}}
+\theta B \gamma \frac{2\ell^4 r^2 r^2 (2r^2-3\beta)+2\gamma^2(2r^2-\beta)^2 + \ell^2 h (12r^6-12 \beta r^4+3\beta^2 r^2-4\beta \gamma^2)}{16 \ell^7 r^6 h^2 {f}^{3/2}}
\bigg]dr \,,\\
  \hat{\omega}^2=&
\bigg[ -\frac{h'}{2}\sqrt{\frac{f}{h}}
+\theta B \frac{4\ell^4 r^2 h^2 (r^2+\beta)+8 \beta \gamma^2 \ell^2 h -(2r^2-\beta)^2 \gamma^2}{16 \ell^6 r^5 h^{3/2} f^{1/2}}
\bigg] dt
 - \theta B \frac{\beta \gamma}{8\ell^3 r^3 \sqrt{fh}}  d\vap \,.
\label{eq:w2_def}
\end{align}

From the vielbeins, one can further construct the deformed metric:
\begin{align}
  ds^2=& -(\hat{e}^0)^2 + (\hat{e}^1)^2 + (\hat{e}^2)^2
\nn\\=&
-\bigg[ h(r) - \theta B\frac{2r^2-\beta}{4\ell^2 r^2} \bigg] dt^2
+ \bigg[\frac{1}{f(r) }
+\theta B \frac{2h(r) \ell^2+2r^2-\beta}{4 \ell^2 r^2 f(r)^2} \bigg] dr^2
+ \bigg[ r^2- \frac{\theta B}{2}  \bigg] d\varphi^2
-\frac{2\gamma}{\ell} dtd\varphi 
\nn\\&
+{\cal O}(\theta^2)
\label{eq:deformed_metric}
 \,.
\end{align}
In the neutral limit $Q\rightarrow 0$ (namely $\beta\rightarrow 0$),
this metric agrees with \eqref{eq:neutral_def_BTZ}.
When one applies the same change of coordinates as in \eqref{eq:coord_change},
the metric recovers the undeformed one
\eqref{eq:crBTZ_sol} with $r$ replaced by $\tilde{r}$.
This implies that the Einstein equation
\begin{align}
  \hat{R}_{\mu\nu} - \frac{1}{2}\hat{g}_{\mu\nu} \hat{R} - \frac{1}{\ell^2} \hat{g}_{\mu\nu}
  =& 8\pi G \hat{\cal T}_{\mu\nu} \,,
\label{eq:Einstein_eq_def1}
\end{align}
is satisfied
if we apply the same coordinate transformation to the right hand side (
the gauge field energy-momentum tensor) simultaneously.

Now we would like to investigate the change of coordinates and the Einstein equation more closely.
The Ricci tensor and the scalar curvature are constructed from the deformed metric
$\hat{g}_{\mu\nu}$ and its Levi-Civita connection, 
\begin{align}
                          \begin{Bmatrix}
                            \rho \\ \mu\nu
                          \end{Bmatrix}
=\frac{1}{2}\hat{g}^{\rho\sigma} \big( \partial_\mu \hat{g}_{\nu\sigma} +\partial_\nu \hat{g}_{\mu\sigma}
-\partial_\sigma \hat{g}_{\mu\nu} \big) \,.
\end{align}
We denote the left hand side of \eqref{eq:Einstein_eq_def1}
as $G_{\mu\nu}^{(\Lambda)}\big( \hat{g},\{ \phantom{a} \} \big)$.
On the other hand, the deformed energy-momentum tensor reads
\begin{align}
  \hat{\cal T}_{\mu\nu}= \frac{1}{\lambda^2} \bigg[ \hat{\mathfrak{f}}_{\mu\rho} \hat{\mathfrak{f}}_{\nu\sigma} \hat{g}^{\rho\sigma}
-\frac{1}{4}\hat{g}_{\mu\nu} \hat{\mathfrak{f}}_{\rho\sigma} \hat{\mathfrak{f}}_{\xi\zeta} \hat{g}^{\rho\xi} \hat{g}^{\sigma\zeta} \bigg] \,,
\label{eq:EM_tensor_def}
\end{align}
where
\begin{align}
  \hat{\mathfrak{f}}_{tr}(r)=&
\mathfrak{f}_{tr}(r) \bigg|_{r \rightarrow r-\frac{\theta B}{4r}}=
\frac{\lambda Q}{r} \bigg(1+\frac{\theta B}{2r^2} \bigg) + {\cal O}(\theta^2) \,,
\label{eq:f_sol_def}
\end{align}
is obtained by applying the inverse of the coordinate transformation \eqref{eq:coord_change}
to the undeformed field strength $\mathfrak{f}_{tr}(r)$.
As a result, two equations of motions are related as follows,
\begin{align}
  G_{\mu\nu}^{(\Lambda)}(g,\{ \phantom{a}\}) = 8\pi G {\cal T}_{\mu\nu}
\quad
\xrightarrow{r\rightarrow r-\frac{\theta B}{4r}}
\quad
G_{\mu\nu}^{(\Lambda)}(\hat{g},\{ \phantom{a}\}) = 8\pi G \hat{\cal T}_{\mu\nu} \,.
\end{align}
It may appear that the deformed metric is again trivial and equivalent to the undeformed one up to a simple coordinate transformation.
However, it turns out that there remains a non-vanishing torsion tensor
in this charged case and the solution
is not related to the undeformed metric just by a coordinate change.
This issue will be discussed in the next subsection.

\subsection{Torsion and Einstein-Cartan gravity}
\label{sec:tors-einst-cart}

The connection in noncommutative space is calculated by use of the deformed vielbeins and
spin connections \eqref{eq:e0_def}--\eqref{eq:w2_def} as follows:
\begin{align}
  \Gamma^\lambda_{\mu\nu}=& \hat{e}^\lambda{}_a \big(\partial_\mu
 \hat{e}_\lambda{}^a + \hat{\omega}_\mu{}^a{}_b \hat{e}_\lambda{}^b \big)
\,.
\end{align}
They are asymmetric with respect to $\mu$ and $\nu$ indices,
and provide nontrivial torsion:
\begin{align}
  T_{\mu\nu}{}^\rho=& \Gamma_{\mu\nu}^\rho - \Gamma_{\nu\mu}^\rho \,.
\end{align}
To be explicit, the non-vanishing components of torsion are
\begin{align}
  T_{tr}{}^0=& -\beta \theta B \frac{r^2 \ell^2 f+\gamma^2}{8\ell^4 r^5 h^{1/2} f} \,,
\qquad
T_{r\vap}{}^0= -\beta \theta B \frac{\gamma}{8 \ell^3 r^3 f h^{1/2}}
\,,\\
  T_{tr}{}^2 =&
-\beta \theta B \frac{\gamma}{4\ell^3 r^4 (fh)^{1/2}} 
\,,\qquad
T_{r\vap}{}^2 =
-\beta \theta B \frac{1}{8\ell^2 r^2 (fh)^{1/2}} 
\label{eq:torsion_values}
\,.
\end{align}
Therefore, the connection associated with the deformed solution is not a Levi-Civita connection
but a more general Affine connection.
Furthermore, the curvature tensors should be calculated by use of $\hat{g}_{\mu\nu}$ and the Affine connection $\Gamma_{\mu\nu}^\rho$,
The non-vanishing components of Einstein tensor $G^{(\Lambda)}_{\mu\nu}(\hat{g},\Gamma)$ (including a cosmological constant
term) are
\begin{align}
  G^{(\Lambda)}_{tt}(\hat{g},\Gamma)=& \beta \frac{\ell^2 r^2 h + 2\gamma^2}{2\ell^4 r^4}
+ \beta \theta B \frac{8\ell^2 r^2 f -2r^4+\beta r^2 + 24 \gamma^2}{16 \ell^4 r^6}
\,,\\
  G^{(\Lambda)}_{t\vap}(\hat{g},\Gamma)=& -\beta \frac{\gamma}{2\ell^3 r^2} - \beta \gamma \theta B\frac{3}{4 \ell^3 r^4}
\,,\\
  G^{(\Lambda)}_{rr}(\hat{g},\Gamma)=& -\beta\frac{1}{2\ell^2 r^2 f} -\beta \theta B \frac{10 \ell^2 r^2 f + 2r^4-\beta r^2 -2\gamma^2}{16 \ell^4 r^6 f^2} 
\,,\\
  G^{(\Lambda)}_{\vap t}(\hat{g},\Gamma)=& -\beta \frac{\gamma}{2\ell^3 r^2} - \beta \gamma \theta B\frac{1}{ \ell^3 r^4}
\,,\\
  G^{(\Lambda)}_{\vap \vap}(\hat{g},\Gamma)=& \beta \frac{1}{2\ell^2} + \beta \theta B \frac{3}{8\ell^2 r^2} \,.
\end{align}
Note that the Einstein tensor $G^{(\Lambda)}_{\mu\nu}(\hat{g},\Gamma)$ is also asymmetric
due to torsion.
Since torsion transforms as a genuine tensor, this solution cannot be related to the undeformed one 
with vanishing torsion by a mere coordinate change.

The theory of gravity with torsion is known as Einstein-Cartan theory of gravitation.
Some features of Einstein-Cartan theory are briefly summarized in the App.\ref{sec:einst-cart-theory}.
As explained there, the equations of motion has an extra contribution depending on torsion,
and they now read
\begin{align}
  G_{\mu\nu}^{(\Lambda)}(\hat{g},\Gamma) -\frac{1}{2}  \overset{*}{\nabla}_\alpha
\big[- \tilde{T}_{\mu\nu}{}^\alpha+\tilde{T}^\alpha{}_{\mu\nu}+\tilde{T}^\alpha{}_{\nu\mu} \big]
= 8\pi G \hat{\cal T}_{\mu\nu} \,,
\label{eq:Einstein-Cartan_eq}
\end{align}
where $\overset{*}{\nabla}_\alpha\equiv \nabla_\alpha + T_\alpha$.  
$T_\alpha$ is the trace of the torsion tensor, while 
$\tilde{T}_{\mu\nu}{}^\rho$ is the deformed one.
Note that \eqref{eq:torsion_values} leads to the vanishing trace of the torsion $T_\alpha=0$
and then $\overset{*}{\nabla}_\alpha= \nabla_\alpha$.

Now we make an interesting observation that the equations of motion \eqref{eq:Einstein-Cartan_eq} are
also satisfied if the deformed energy momentum tensor \eqref{eq:EM_tensor_def} is adopted.
Namely, we have confirmed the following equivalence under the change of coordinate:
\begin{align}
  G_{\mu\nu}^{(\Lambda)}(g,\{ \phantom{a}\})
\quad
\xrightarrow{r\rightarrow r-\frac{\theta B}{4r}}
\quad
&
G_{\mu\nu}^{(\Lambda)}(\hat{g},\{ \phantom{a}\}) 
\nn\\=&
  G_{\mu\nu}^{(\Lambda)}(\hat{g},\Gamma) -\frac{1}{2}  \overset{*}{\nabla}_\alpha
\big[- \tilde{T}_{\mu\nu}{}^\alpha+\tilde{T}^\alpha{}_{\mu\nu}+\tilde{T}^\alpha{}_{\nu\mu} \big]
\,.
\end{align}
In other words, the effect of torsion at the left hand side of equation \eqref{eq:Einstein-Cartan_eq} appears to cancel out.

So far we have observed a part of the set of equations of motion.
In Einstein-Cartan theory, there are also equations of motion from the variation
with respect to the torsion:
\begin{align}
  K_\rho{}^{\nu\mu}+T^\nu\delta^\mu_\rho-T^\mu\delta^\nu_\rho
=& -\frac{\delta I_\text{gauge}}{\delta T_{\mu\nu}{}^\rho} \,,
\end{align}
where $K_\rho{}^{\nu\mu}$ is the contortion.
The standard action for deformed $U(1)$ gauge field action does not couple to torsion, therefore the right hand side is zero.
Since the left hand side is nonvanishing, the matter part of the
action should also be modified such that it couples to torsion.
In the next section, we will treat these two equations of motions in a unified way by use of Chern-Simons equations of
motion.

\subsection{The Chern-Simons equations of motion and the matter energy-momentum tensor}
\label{sec:matt-energy-moment}

After the deformation, we may assume that the matter part action is replaced as
\begin{align}
  \hat{I}_\text{gauge}=& 
-\frac{1}{\lambda^2} \int d^3 x\, \sqrt{-\hat{g}} \hat{\mathfrak{f}}_{\mu\nu} \hat{\mathfrak{f}}_{\rho\sigma} \hat{g}^{\mu\rho} \hat{g}^{\nu\sigma}
+ I'_\text{gauge} \,,
\end{align}
where $I'_\text{gauge}$ includes the coupling to $B_\mu^{(\pm)}$ and torsion (or spin connection),
and should vanish in the commutative limit $\theta \rightarrow 0$
(or to provide decoupled equations of motion).
The generic form is argued in the Sec.~\ref{sec:coupl-matt-fields}.
We will calculate the energy-momentum tensor and the spin density tensor from this action,
and choose the coupling constant to determine the correction term.
In terms of Chern-Simons gauge fields,
the correction term should satisfy the following equations of motion,
\begin{align}
    \frac{\delta I_{CS}}{\delta \hat{A}_\mu^{(\pm)a}}=&
\mp \frac{k}{4\pi} \epsilon^{\mu\rho\sigma}
\bigg[
\eta_{ab}\partial_\rho \hat{A}_\sigma^{(\pm)b}
-\frac{\epsilon_{abc}}{2}\hat{A}_\rho^{(\pm)b}  \star \hat{A}_\sigma^{(\pm)c}
+\frac{i}{6}\eta_{ab}\big( \hat{A}_\rho^{(\pm)b} \star \hat{B}_\sigma^{(\pm)} +
                                               \hat{B}_\rho^{(\pm)} \star \hat{A}_\sigma^{(\pm)b} \big)
 \bigg]
\nn\\=& -\frac{\delta \hat{I}_\text{gauge}}{\delta \hat{A}_\mu^{(\pm)a}} 
\label{eq:eqm_A_def0}
\,,\\
  \frac{\delta I_{CS}}{\delta \hat{B}_\mu^{(\pm)}}=&
\pm \frac{k}{4\pi} \epsilon^{\mu\rho\sigma} 
\bigg[
\partial_\rho \hat{B}_\sigma^{(\pm)}
-\frac{i}{6} \eta_{ab} \hat{A}_\rho^{(\pm)a} \star \hat{A}_\sigma^{(\pm)b}
+ \frac{i}{2}\hat{B}_\rho^{(\pm)} \star \hat{B}_\sigma^{(\pm)}
\bigg]=
-\frac{\delta \hat{I}_\text{gauge}}{\delta \hat{B}_\mu^{(\pm)}}
\label{eq:eqm_B_def0} \,,\\
\frac{\delta \hat{I}_\text{gauge}}{\delta \hat{\mathfrak{a}}_\mu}
=&  
\frac{1}{\lambda^2} \nabla_\rho \hat{\mathfrak{f}}^{\rho\mu} + \frac{\delta {I}'_\text{gauge}}{\delta \hat{\mathfrak{a}}_\mu}
=0\,.
\label{eq:eom_a}
\end{align}
As for $\hat{A}_\mu^{(\pm)a}$ and $\hat{B}_\mu^{(\pm)}$, we consider the explicit solution via the Seiberg-Witten map.
We first require that the last equation \eqref{eq:eom_a} leads to the solution
$\hat{\mathfrak{f}}_{\mu\nu}$ in \eqref{eq:f_sol_def}.
We also assume that all the fields are functions of only $r$ so we can replace $\star$ product with a usual product and those terms for interaction between $\hat{A}_\mu^{(\pm)a}$
and $\hat{B}_\mu^{(\pm)}$ are dropped.
At the end, \eqref{eq:eqm_A_def0} becomes
\begin{align}
    \frac{\delta \hat{I}_{CS}}{\delta \hat{A}_\mu^{(\pm)a}} =&
\mp \frac{k}{4\pi} \epsilon^{\mu\rho\sigma} \eta_{ab}
\bigg[ \partial_\rho \hat{A}_\sigma^{(\pm)b} - \frac{\epsilon^b{}_{cd}}{2} \hat{A}_\rho^{(\pm)c} \hat{A}_\sigma^{(\pm)d} \bigg]
\nn\\=&
\mp \frac{k}{4\pi} \epsilon^{\mu\rho\sigma} \eta_{ab}
\bigg[
\partial_\rho \hat\omega_\sigma{}^b
-\frac{\epsilon^b{}_{cd}}{2}\bigg( \hat\omega_\rho{}^c \hat\omega_\sigma{}^d + \frac{1}{\ell^2}\hat{e}_\rho{}^c \hat{e}_\sigma{}^d \bigg)
\pm \frac{1}{\ell} \big(\partial_\rho \hat{e}_\sigma{}^b - \epsilon^b{}_{cd} \hat{e}_\rho{}^c \hat\omega_\sigma{}^d \big) \bigg]
\nn\\=&
\pm \frac{k}{4\pi} g^{\mu\zeta} e^\delta{}_a \sqrt{-g} G^{(\Lambda)}_{\zeta\delta}(\hat{g},\Gamma)
- \frac{k}{8\pi \ell} \epsilon^{\mu\rho\sigma} \eta_{ab} T_{\rho\sigma}{}^b
\,.
\end{align}
It is easy to check that only the first term survives in \eqref{eq:eqm_B_def0}.
Therefore the equations of motion become
\begin{align}
  \pm \frac{k}{4\pi} \hat{g}^{\mu\zeta} \hat{e}^\delta{}_a \sqrt{-\hat{g}} 
G^{(\Lambda)}_{\zeta\delta}(\hat{g},\Gamma)
- \frac{k}{8\pi \ell} \epsilon^{\mu\rho\sigma} \eta_{ab} T_{\rho\sigma}{}^b =&
\mp \frac{\ell}{2} \eta_{ac} \cdot \sqrt{-\hat{g}} \hat{{\cal T}}^{\mu\rho} \hat{e}_\rho^c
-\frac{\delta I'_\text{gauge}}{\delta \hat{A}_\mu^{(\pm)a}} \,,\\
\pm \frac{k}{4\pi} \epsilon^{\mu\rho\sigma} 
\partial_\rho \hat{B}_\sigma^{(\pm)}
=& - \frac{\delta {I}'_\text{gauge}}{\delta \hat{B}_\mu^{(\pm)}} \,.
\end{align}
Or equivalently, one may write it as
\begin{align}
  G^{(\Lambda)}_{\mu\nu}(\hat{g},\Gamma) =&
8\pi G \hat{{\cal T}}_{\mu\nu}
-\frac{2\pi}{k} \hat{g}_{\mu\zeta} \hat{e}_\nu{}^a
\bigg(
\frac{\delta I'_\text{gauge}}{\delta \hat{A}_\zeta^{(+)a}}
-
\frac{\delta I'_\text{gauge}}{\delta \hat{A}_\zeta^{(-)a}} \bigg) \,,\\
T_{\mu\nu}{}^a=& 8\pi G \eta^{ab} \epsilon_{\mu\nu\zeta}
\bigg(
\frac{\delta I'_\text{gauge}}{\delta \hat{A}_\zeta^{(+)b}}
+
\frac{\delta I'_\text{gauge}}{\delta \hat{A}_\zeta^{(-)b}} \bigg) \,.
\end{align}
By comparing with \eqref{eq:Einstein-Cartan_eq}, the first equation implies that
\begin{align}
\frac{2\pi}{k} \hat{g}_{\mu\zeta} \hat{e}_\nu{}^a
\bigg(
\frac{\delta I'_\text{gauge}}{\delta \hat{A}_\zeta^{(+)a}}
-
\frac{\delta I'_\text{gauge}}{\delta \hat{A}_\zeta^{(-)a}} \bigg)
=&
-\frac{1}{2}  \overset{*}{\nabla}_\alpha
\big[- \tilde{T}_{\mu\nu}{}^\alpha+\tilde{T}^\alpha{}_{\mu\nu}+\tilde{T}^\alpha{}_{\nu\mu} \big]
\,.
\end{align}
We therefore find a set of conditions for the correction term in the matter part as follows:
\begin{align}
  \frac{\delta I'_\text{gauge}}{\delta \hat{A}_\mu^{(\pm)a}} =&
\pm \frac{\ell}{16 \pi G} \hat{g}^{\mu\nu} \hat{e}^\rho{}_a
\overset{*}{\nabla}_\alpha
\big[- \tilde{T}_{\nu\rho}{}^\alpha+\tilde{T}^\alpha{}_{\nu\rho}+\tilde{T}^\alpha{}_{\rho\nu} \big]
-\frac{1}{32\pi G}\epsilon^{\mu\nu\rho} \eta_{ab} T_{\nu\rho}{}^b \,,\\
\frac{\delta {I}'_\text{gauge}}{\delta \hat{B}_\mu^{(\pm)}}
=& \pm \frac{\ell}{16\pi G} \epsilon^{\mu\rho\sigma}  \partial_\rho \hat{B}_\sigma^{(\pm)} \,,
\label{eq:I_matter_B_cond}\\
\frac{\delta {I}'_\text{gauge}}{\delta \hat{\mathfrak{a}}_\mu}
=& -\frac{1}{\lambda^2} \nabla_\rho \hat{\mathfrak{f}}^{\rho\mu}  \,.
\end{align}
We have not fixed the explicit form of the correction term due to its complexity.
Here we simply present the necessary conditions for the $\theta$ dependent correction term
for the matter part of the action.\footnote{%
By use of the change of the variables \eqref{eq:variation_change}, one can also consider
these relations in terms of the variations with respect to the metric and the torsion.}

\subsection{Coordinate change}
\label{sec:coordinate-change}

Finally, we briefly comment on the change of coordinates
\eqref{eq:coord_change}. 
There is a subtle point on the regions that the radial
coordinate covers.
Recall that $\tilde{r}$ covers $0\leq \tilde{r} < \infty$ and it has one-to-one correspondence to
the region $\sqrt{\theta B}/2 \leq r <\infty$.
Except for the vicinity of center in the deformed geometry, $0\leq r \leq \sqrt{\theta B}/2$, 
it can be mapped to the undeformed geometry.
Now we investigate the angular part of the metric, namely $\tilde{r}^2 d\vap^2$ or $\big( r^2 - \frac{\theta B}{2}\big) d\vap^2$.
In the deformed metric, the radial coordinate $r$ makes sense only for the region $r \geq \sqrt{\theta B/2}$.
Therefore, in the deformed geometry, there appears an effective minimum length scale
$r_\text{min}=\sqrt{\theta B/2}$.
This may not be so surprising; in the current formulation, the
noncommutative parameter appears only in the combination of $\theta B$
and we learn that $\sqrt{\theta B}$ serves a characteristic length scale in the
noncommutative geometry.
Now let $r_+$ be the location of the horizon of noncommutative BTZ measured in $r$ coordinate, that is, the largest root of $\hat{g}_{tt}(r_+)=0$.
As long as $r_+\geq r_\text{min}$, we can see the correspondence to
the undeformed BTZ solution.
On the other hand, a black hole of the size $r_+ < r_\text{min}$ is not
well-defined in the noncommutative side.

Finally, we comment on a subtle issue on the coordinate invariance of noncommutative gravity.\footnote{%
We thank an anonymous referee to raise this point.}
The action of noncommutative gravity, by using Chern-Simons formulation, 
is invariant under a deformed coordinate transformation, which is reduced to the usual diffeomorphism
in the commutative limit \cite{Cacciatori:2002gq}.
As studied in Ref.\cite{ChangYoung:2009sn}, two noncommutative theories obtained by the Seiberg-Witten map
with the rectangular and polar coordinates are distinct; this implies that we cannot map one to the other
by a simple coordinate change in the noncommutative theory.
In the present case, the situation is much simpler; the change is only for the radial coordinate $r$
and it does not change the noncommutative algebra
(unlike the change between the rectangular and polar coordinates discussed in Ref.\cite{ChangYoung:2009sn}),
On top of that, all the relevant function in the metric depends only on $r$.
In other words, the noncommutativity is irrelevant when we consider a deformed coordinate transform (star product
simply reduces to ordinary product), and the transformed solution satisfies the conventional Einstein equation.
Together with the uniqueness of local solution for the vacuum case, 
this could be the very reason why 
 the deformed solution is related
to the undeformed one by the coordinate change \eqref{eq:coord_change}.
In the case of Einstein-Maxwell theory, since solutions do not have to be unique and
a noncommutative extension involves nontrivial torsion, we cannot see immediately
why the simple relation still holds.
It may be related to the fact that since the functions in the commutative solution
only involve the radial coordinate $r$, it could be sufficient to consider a
commutative version of Einstein-Cartan theory while we only concern the
equation of motion.  
We expect a simple coordinate change like \eqref{eq:coord_change}
becomes impossible for a generic geometry whose metric functions
depend on both $r$ and $\vap$ coordinates. 
We, however, leave this complexity for future studies.

\section{Conclusion}
\label{sec:conclusion}

In this paper, we have explored the charged rotating BTZ black hole
geometry by use of Chern-Simons formulation in $2+1$ dimensional 
gravity and the Seiberg-Witten map.

The noncommutativity in question is the one between the radial coordinate and the angular coordinate, namely $[r^2 , \vap]=2i\theta$.
The noncommutative deformation for the pure gravity part is introduced by the Seiberg-Witten map
for the Chern-Simons gauge fields where two extra $U(1)$ gauge fields are added.
The deformation for the matter gauge field part is to be determined to satisfy the deformed equations of motion.

It is found that as with the neutral case, the deformed metric 
is related to the undeformed one via a simple coordinate transformation.
Through this observation, we discover that the deformation of 
the matter energy-momentum tensor can also be obtained by the same coordinate change.
Nevertheless, there appears nonvanishing torsion that is proportional to the noncommutativity parameter and 
it cannot be eliminated by a coordinate change. 
We thus analyze the equations of motion in the framework of Einstein-Cartan torsion gravity.
It is found that with the same deformed matter energy-momentum tensor, the equations of motion
derived from torsion gravity are also satisfied.
Though we have not yet fixed the action of deformed matter completely, a set of conditions
for the correction term are presented.

There are several issues to be clarified.
Firstly, we do not fully understand why the noncommutative deformation is represented by a simple change of 
the radial coordinate.
One can verify that the result of the Seiberg-Witten map for the difference of the gauge fields, say 
$A_\mu^{(+)a}-A_\mu^{(-)a}$, can be obtained by the same coordinate change, however the sum is not.
Therefore, the deformed vielbeins $\hat{e}_\mu{}^a$ are related to the undeformed ones via the coordinate change,
however the spin connections are not.
This subtle difference leads to the nontrivial torsion in the deformed background.
One may argue that the gauge field representation has some nonphyiscal degrees of freedom, but the appearance of torsion is
physical and cannot be trivially eliminated by the coordinate change.

We observe that the matter gauge field couples to $B^{(\pm)}$ after the deformation.
In Ref.\cite{Cacciatori:2002gq}, the authors argued that in the noncommutative
Chern-Simons theory there is a coupling term between $(B^{(+)}+B^{(-)})$
and torsion.  One thus may guess that the coupling to torsion appears via
$B^{(\pm)}$.  We remark that the noncommutative extension of
torsion constructed in Ref.\cite{Cacciatori:2002gq} cosnsists of two parts: a standard part (which we call torsion
in this paper) and a term like $\eta_{ab}(\omega^a \swedge e^b + e^a \swedge
\omega^b)$ (which trivially vanishes in the commutative limit).  However, $B^{(\pm)}$ couples only to the latter.
Therefore, the coupling between the geometrical part of torsion to the other
degrees of freedom remains unclear.

Secondly, the admitted minimal black holes discussed in the Sec. \ref{sec:coordinate-change} may imply 
that the noncommutative space-time has its own entropy, i.e. $S \propto \theta B$ in a region of Planckian size, presuming the area law still applies.
This reminds us of the spin foam model in the loop quantum gravity \cite{Reisenberger:1996pu} and we wonder if $e^S$ counts the spin combination.

Thirdly, it is curious which properties of the charged BTZ black hole are changed or unchanged after the deformation.
The torsion may affect the property of black holes through the change of metric \cite{torsion_BH}.
However, in our case as long as we look at the metric only, we do not see the difference.
It is interesting to see whether this is a peculiar feature of the current solution, or this may happen in
a broader setup of $2+1$ dimensional gravity with noncommutativity.
On top of that, it should also be important to fix the deformation of the matter part action and 
examine how the matter part action couples to the torsion or the extra $U(1)$ gauge fields
$B_\mu^{(\pm)}$.

Finally, we would like to comment on results made in Ref.\cite{Mukherjee:2006nd, Banerjee:2007th}, where noncommutative structures, including Lie algebraic structure ones, are considered in $3+1$ dimensional gravity.  It was argued that the first order correction vanishes under the condition
of vanishing classical torsion.
In our construction, however, we include a matter field whose deformation is not completely fixed
by the Seiberg-Witten map and the deformed solution has nontrivial torsion.
Therefore, our result would not be immediately contradict to their results.
Since there appears a simple relation between the deformed geometry with torsion
and the undeformed one, it is interesting to investigate the applicability of our argument
to generic backgrounds in torsion gravity.

\textbf{Note added:} Upon completing this work, there appeared a paper \cite{Ciric:2016isg}
, which considered a noncommutative deformation in four dimensional gravity.  
They also observed the emergence of torsion.

\section*{Acknowledgment}
\label{sec:acknowledement}
This work is supported in parts by the Taiwan's Ministry of Science and Technology (grant No. 102-2112-M-033-003-MY4) and the National Center for Theoretical Science.

\appendix

\section{Conventions and notations}
\label{sec:conventions}

We summarize our conventions and notations in this paper here.

\subsection{Seiberg-Witten map}
\label{sec:seiberg-witten-map}

Seiberg and Witten showed that a field theory on D-branes with a background $B$ field
can be formulated as a conventional Yang-Mills theory or a noncommutative Yang-Mills theory
depending on the regulator we choose, Pauli-Villars or point-splitting respectively \cite{Seiberg:1999vs}.
The gauge transformation is now defined by use of Moyal product as
\begin{align}
  \hat{\delta}_{\hat{\xi}} \hat{A}_\mu =&
\partial_\mu \hat{\xi} -  \hat{\xi} \star \tilde{A}_\mu + \hat{A}_\mu \star \hat{\xi}
\nn\\=&
\partial_\mu \hat{\xi} 
-\frac{i}{2} \theta^{\nu\rho} \big( \partial_\nu \hat{\xi} \partial_\rho \hat{A}_\mu
-\partial_\nu \hat{A}_\mu \partial_\rho \hat{\xi}  \big)
+{\cal O}(\theta^2) 
 \,.
\end{align}
The Seiberg-Witten map is defined as a compatibility condition of
gauge transformation and a mapping between $A$ and $\hat{A}$,
\begin{align}
  \hat{A}(A)+ \hat{\delta}_{\hat{\xi}} \hat{A}(A) =
\hat{A}(A+\delta_\xi A) \,,
\label{eq:SW_gaugeRel}
\end{align}
for infinitesimal $\xi$ and $\hat{\xi}$.
The solution is
\begin{align}
  \hat{A}_\mu(A)=& A_\mu - \frac{i }{4}\theta^{\nu\rho}  \{ A_\nu, \partial_\rho A_\mu + F_{\rho\mu} \}
+{\cal O}(\theta^2)
\label{eq:SWmap1}
 \,,\\
\hat{\xi}(\xi,A)=& \xi + \frac{i }{4}\theta^{\mu\nu} \{ \partial_\mu \xi , A_\nu \}
+ {\cal O}(\theta^2) \,,
\end{align}
where $\{ f, g\}=fg + gf$ is the anti-commutator with respect to the conventional matrix product.

\subsection{Some notations and $U(1,1)$ generators}

The epsilon tensor is $\epsilon_{012}=-\epsilon^{012}=1$.
We define for a spin connection 1-form $\omega^a{}_b$,
\begin{align}
  \omega_a= - \frac{1}{2}\epsilon_{abc} \omega^{bc} \,.
\end{align}

For the neutral BTZ black holes, $r_\pm$ is defined by
\begin{align}
  r_\pm^2 =& 4 G \ell^2 \bigg( M \pm \sqrt{M^2-\frac{J^2}{\ell^2} }\bigg) \,,\\
M=& \frac{r_+^2+r_-^2}{8G \ell^2} \,,
\qquad
J= \frac{r_+r_-}{4G \ell} \,.
\end{align}

Our convention of $U(1,1)$ generators is
\begin{align}
  \tau_0 =& \frac{i}{2}\sigma_3 
\,, \quad
  \tau_1 = \frac{1}{2}\sigma_1 
 \,,\quad
   \tau_2 = \frac{1}{2}\sigma_2 
\,,\quad
  \tau_3 = \frac{i}{2}\bm{1}_2 
\,,
\label{eq:SL22R_gen}
\end{align}
with $a,b=0,1,2$, $A,B=0,1,2,3$, $\eta_{AB}=\text{diag}(-1,1,1,-1)$
and they satisfy
\begin{align}
&
  g_{AB} = \tr(\tau_A\tau_B) = \frac{1}{2}\eta_{AB} \,,
\qquad
[\tau_A,\tau_B] = - \epsilon_{AB}{}^C \tau_C \,,
\quad
\epsilon_{AB}{}^C=
  \begin{cases}
    \epsilon_{ab}{}^c \\ \epsilon_{ab}{}^3=\epsilon_{3a}{}^b=0
  \end{cases}
\,,\\&
\{ \tau_a, \tau_b \} = \frac{1}{2} \eta_{ab}\mathbf{1}_2 \,,
\quad
\{ \tau_A, \tau_3 \} = i \tau_A \,,
\qquad
\tr \, (\tau_a\tau_b\tau_c) = - \frac{1}{4} \epsilon_{abc} \,,
\quad
\tr \, (\tau_a \tau_b \tau_3)= \frac{i}{4} \eta_{ab} \,.
\end{align}

By use of the chain rule, we can convert the variation with respect to the gauge fields
to those with respect to the metric and the torsion as
\begin{align}
  \frac{\delta}{\delta A_\mu^{(\pm)a}}=&
\mp \frac{\ell}{2}
\bigg[
2 g^{\mu\alpha} e^\beta{}_a  \frac{\delta}{\delta g^{\alpha\beta}}
+e^\beta{}_a
\big[ 
\delta^\alpha_\beta T_{\rho\sigma}{}^\mu
+\delta^\mu_\rho  \Gamma_{\sigma\beta}^\alpha
-\delta^\mu_\sigma   \Gamma_{\rho\beta}^\alpha
\big] \frac{\delta}{\delta T_{\rho\sigma}{}^\alpha}
\bigg]
\nn\\&
+\frac{1}{2} \epsilon^b{}_{ca} e^\alpha{}_b \big( \delta^\mu_\rho e_\sigma{}^c 
-\delta^\mu_\sigma e_\rho{}^c
\big)
\frac{\delta}{\delta T_{\rho\sigma}{}^\alpha}
 \,.
\label{eq:variation_change}
\end{align}

\subsection{Einstein-Cartan theory of torsion gravity}
\label{sec:einst-cart-theory}

The Einstein-Cartan theory of gravitation is a generalization of
Einstein's theory of general relativity to allow torsion in
space-time.  It can be regarded as a gauge theory of the Poincar{\'e}
symmetry instead of the Lorentz symmetry\cite{Kibble:1961ba}.  While
curvature is related to the energy momentum tensor with Lorentz
symmetry, torsion is related to the density of intrinsic angular
momentum or spin.
For some overview of torsion gravity, see Ref.\cite{Torsion_gravity}.

The vielbeins $e_\mu{}^a$ relate to the metric
by $g_{\mu\nu}=e_\mu{}^a \epsilon_\nu{}^b \eta_{ab}$, where $\eta_{ab}=\text{diag} (-1,1,1)$,
and its inverse is $e_\mu{}^a e^\nu{}_a=\delta^\nu_\mu$ and
$e_\mu{}^a e^\mu{}_b=\delta^a_b$.
With spin connections $\omega_\mu{}^a{}_b$,
Affine connections are defined by
\begin{align}
  \Gamma^\nu_{\mu\lambda}=& e^\nu{}_a \big( \partial_\mu e_\lambda{}^a + \omega_\mu{}^a{}_b e_\lambda{}^b\big) \,,
\end{align}
and the torsion tensor is
\begin{align}
  T_{\mu\nu}{}^a =& 
\partial_\mu {e}_\nu{}^a
-\partial_\nu {e}_\mu{}^a
+\hat{\omega}_\mu{}^a{}_b {e}_\nu{}^b
-\hat{\omega}_\nu{}^a{}_b {e}_\mu{}^b \,.
\end{align}
The curvature tensor is
\begin{align}
    R^{\lambda}{}_{\rho\mu\nu} =&
\partial_\mu \Gamma^\lambda_{\nu\rho} -\partial_\nu \Gamma^\lambda_{\mu\rho} 
+ \Gamma^\lambda_{\mu\xi} \Gamma^\xi_{\nu\rho} -  \Gamma^\lambda_{\nu\xi} \Gamma^\xi_{\mu\rho}
\,,
\end{align}
and the Ricci tensor and the scalar curvature are defined by
$R_{\mu\nu}=R^\rho{}_{\mu\rho\nu}$ and $R=g^{\mu\nu}R_{\mu\nu}$ respectively.

In Einstein-Cartan theory of torsion gravity, the metric $g_{\mu\nu}$
and the connection $\Gamma_{\mu\nu}^\rho$ are treated as independent variables.
When we consider the equations of motion, we can take the variation of torsion tensor
instead of the connection.
The action is given by the usual Einstein-Hilbert form,
\begin{align}
  I_G= \frac{1}{16 \pi G} \int d^3 x \, \sqrt{-g} R \,,
\end{align}
and its variations give
\begin{align}
\frac{16\pi G}{\sqrt{-g}} \frac{\delta I_G}{\delta g^{\mu\nu}}=&
  G_{\mu\nu}^{(\Lambda)}(g,\Gamma) -\frac{1}{2}  \overset{*}{\nabla}_\alpha
\big[- \tilde{T}_{\mu\nu}{}^\alpha+\tilde{T}^\alpha{}_{\mu\nu}+\tilde{T}^\alpha{}_{\nu\mu} \big] \,,\\
16\pi G \frac{\delta I_G}{\delta T_{\mu\nu}{}^\rho}=&
K_\rho{}^{\nu\mu}+T^\nu\delta^\mu_\rho-T^\mu\delta^\nu_\rho \,,
\end{align}
where $\overset{*}{\nabla}_\alpha \equiv \nabla_\alpha + T_\alpha$
with $\nabla_\alpha$ being a covariant derivative 
and $T_\alpha$ the trace of the torsion tensor
$T_{\alpha\nu}{}^a \hat{e}^\nu{}_a$.
The contorsion tensor $K_{\mu\nu\sigma}$ is defined as
\begin{align}
  K_{\mu\nu\sigma} = \frac{1}{2} \big(T_{\mu\nu\sigma} - T_{\nu\sigma\mu} + T_{\sigma\mu\nu}  \big) \,,
\end{align}
and
$\tilde{T}_{\mu\nu}{}^\rho$ is known as the deformed torsion tensor:
\begin{align}
  \tilde{T}_{\mu\nu}{}^\rho= T_{\mu\nu}{}^\rho + \delta^\rho_\mu T_\nu -
  \delta^\rho_\nu T_\mu 
\,.
\end{align}

Finally, by use of the one forms $e^a=e_\mu{}^a dx^\mu$ and $\omega^a{}_b=\omega_\mu{}^a{}_b dx^\mu$,
the torsion and the curvature two forms are written as 
\begin{align}
  T^a=& De^a=de^a + \omega^a{}_b \wedge e^b \,,\\
R^a{}_b =& d\omega^a{}_b + \omega^a{}_c \wedge \omega^c{}_b \,.
\end{align}



\begin{thebibliography}{99}

\bibitem{NC_geom}
For example,
  H.~S.~Snyder,
  ``Quantized space-time,''
  Phys.\ Rev.\  {\bf 71} (1947) 38.
  doi:10.1103/PhysRev.71.38;\\
  A.~Connes, M.~R.~Douglas and A.~S.~Schwarz,
  ``Noncommutative geometry and matrix theory: Compactification on tori,''
  JHEP {\bf 9802} (1998) 003
  doi:10.1088/1126-6708/1998/02/003
  [hep-th/9711162];\\
  J.~Madore,
  ``The Fuzzy sphere,''
  Class.\ Quant.\ Grav.\  {\bf 9} (1992) 69.
  doi:10.1088/0264-9381/9/1/008

\bibitem{Minwalla:1999px}
  S.~Minwalla, M.~Van Raamsdonk and N.~Seiberg,
  ``Noncommutative perturbative dynamics,''
  JHEP {\bf 0002} (2000) 020
  doi:10.1088/1126-6708/2000/02/020
  [hep-th/9912072].

\bibitem{Yoneya:2000bt} 
  T.~Yoneya,
  ``String theory and space-time uncertainty principle,''
  Prog.\ Theor.\ Phys.\  {\bf 103}, 1081 (2000)
  doi:10.1143/PTP.103.1081
  [hep-th/0004074].



\bibitem{NC_gravity}
For example,
  E.~Harikumar and V.~O.~Rivelles,
``Noncommutative Gravity,''
  Class.\ Quant.\ Grav.\  {\bf 23}, 7551 (2006)
  doi:10.1088/0264-9381/23/24/024
  [hep-th/0607115].\\
  G.~Fucci and I.~G.~Avramidi,
  ``Noncommutative Einstein Equations,''
  Class.\ Quant.\ Grav.\  {\bf 25}, 025005 (2008)
  doi:10.1088/0264-9381/25/2/025005
  [arXiv:0709.0015 [gr-qc]].\\
  M.~Chaichian, M.~Oksanen, A.~Tureanu and G.~Zet,
  ``Gauging the twisted Poincare symmetry as noncommutative theory of gravitation,''
  Phys.\ Rev.\ D {\bf 79}, 044016 (2009)
  doi:10.1103/PhysRevD.79.044016
  [arXiv:0807.0733 [hep-th]].\\
  M.~Kober,
  ``Canonical Noncommutativity Algebra for the Tetrad Field in General Relativity,''
  Class.\ Quant.\ Grav.\  {\bf 28}, 225021 (2011)
  doi:10.1088/0264-9381/28/22/225021
  [arXiv:1107.1071 [hep-th]].\\
  M.~Kober,
  ``Canonical quantum gravity on noncommutative space-time,''
  Int.\ J.\ Mod.\ Phys.\ A {\bf 30}, no. 17, 1550085 (2015)
  doi:10.1142/S0217751X15500852
  [arXiv:1409.1751 [gr-qc]].


\bibitem{Dolan:2006hv} 
  B.~P.~Dolan, K.~S.~Gupta and A.~Stern,
  ``Noncommutative BTZ black hole and discrete time,''
  Class.\ Quant.\ Grav.\  {\bf 24}, 1647 (2007)
  doi:10.1088/0264-9381/24/6/017
  [hep-th/0611233].
\bibitem{Spallucci:2009zz} 
  E.~Spallucci, A.~Smailagic and P.~Nicolini,
  ``Non-commutative geometry inspired higher-dimensional charged black holes,''
  Phys.\ Lett.\ B {\bf 670}, 449 (2009)
  doi:10.1016/j.physletb.2008.11.030
  [arXiv:0801.3519 [hep-th]].

\bibitem{CS_grav}
  A.~Achucarro and P.~K.~Townsend,
  ``A Chern-Simons Action for Three-Dimensional anti-De Sitter Supergravity Theories,''
  Phys.\ Lett.\ B {\bf 180} (1986) 89;\\
  E.~Witten,
  ``(2+1)-Dimensional Gravity as an Exactly Soluble System,''
  Nucl.\ Phys.\ B {\bf 311} (1988) 46.

\bibitem{Mukherjee:2006nd}
  P.~Mukherjee and A.~Saha,
  ``A Note on the noncommutative correction to gravity,''
  Phys.\ Rev.\ D {\bf 74} (2006) 027702
  doi:10.1103/PhysRevD.74.027702
  [hep-th/0605287].


\bibitem{Banerjee:2007th} 
  R.~Banerjee, P.~Mukherjee and S.~Samanta,
  ``Lie algebraic noncommutative gravity,''
  Phys.\ Rev.\ D {\bf 75}, 125020 (2007)
  doi:10.1103/PhysRevD.75.125020
  [hep-th/0703128].
\bibitem{Rivelles:2013ica} 
  V.~O.~Rivelles,
  ``Ambiguities in the Seiberg-Witten map and emergent gravity,''
  Class.\ Quant.\ Grav.\  {\bf 31}, 025011 (2013)
  doi:10.1088/0264-9381/31/2/025011
  [arXiv:1304.5483 [hep-th]].

\bibitem{Pinzul:2005ta} 
  A.~Pinzul and A.~Stern,
  ``Noncommutative AdS**3 with quantized cosmological constant,''
  Class.\ Quant.\ Grav.\  {\bf 23}, 1009 (2006)
  doi:10.1088/0264-9381/23/3/024
  [hep-th/0511071].


\bibitem{Banados:1992wn} 
  M.~Banados, C.~Teitelboim and J.~Zanelli,
  ``The Black hole in three-dimensional space-time,''
  Phys.\ Rev.\ Lett.\  {\bf 69}, 1849 (1992)
  doi:10.1103/PhysRevLett.69.1849
  [hep-th/9204099].

\bibitem{Clement}
  G.~Clement,
  ``Classical solutions in three-dimensional Einstein-Maxwell cosmological gravity,''
  Class.\ Quant.\ Grav.\  {\bf 10} (1993) L49.
  doi:10.1088/0264-9381/10/5/002;\\
  G.~Clement,
  ``Spinning charged BTZ black holes and selfdual particle - like solutions,''
  Phys.\ Lett.\ B {\bf 367} (1996) 70
  doi:10.1016/0370-2693(95)01464-0
  [gr-qc/9510025].


\bibitem{Martinez:1999qi}
  C.~Martinez, C.~Teitelboim and J.~Zanelli,
  ``Charged rotating black hole in three space-time dimensions,''
  Phys.\ Rev.\ D {\bf 61} (2000) 104013
  doi:10.1103/PhysRevD.61.104013
  [hep-th/9912259].

\bibitem{Banados:2001xw}
  M.~Banados, O.~Chandia, N.~E.~Grandi, F.~A.~Schaposnik and G.~A.~Silva,
  ``Three-dimensional noncommutative gravity,''
  Phys.\ Rev.\ D {\bf 64} (2001) 084012

\bibitem{ChangYoung:2008my}
  E.~Chang-Young, D.~Lee and Y.~Lee,
 ``Noncommutative BTZ Black Hole in Polar Coordinates,''
  Class.\ Quant.\ Grav.\  {\bf 26} (2009) 185001
  doi:10.1088/0264-9381/26/18/185001
  [arXiv:0808.2330 [hep-th]].

\bibitem{timelikeNC}
For example,
  N.~Seiberg, L.~Susskind and N.~Toumbas,
  ``Space-time noncommutativity and causality,''
  JHEP {\bf 0006} (2000) 044
  [hep-th/0005015];\\
  J.~Gomis and T.~Mehen,
  ``Space-time noncommutative field theories and unitarity,''
  Nucl.\ Phys.\ B {\bf 591} (2000) 265
  doi:10.1016/S0550-3213(00)00525-3
  [hep-th/0005129].


\bibitem{Cacciatori:2002gq}
  S.~Cacciatori, D.~Klemm, L.~Martucci and D.~Zanon,
  ``Noncommutative Einstein-AdS gravity in three-dimensions,''
  Phys.\ Lett.\ B {\bf 536} (2002) 101


\bibitem{Seiberg:1999vs}
  N.~Seiberg and E.~Witten,
  ``String theory and noncommutative geometry,''
  JHEP {\bf 9909} (1999) 032
  doi:10.1088/1126-6708/1999/09/032
  [hep-th/9908142].

\bibitem{Grandi:2000av}
  N.~E.~Grandi and G.~A.~Silva,
  ``Chern-Simons action in noncommutative space,''
  Phys.\ Lett.\ B {\bf 507} (2001) 345
  doi:10.1016/S0370-2693(01)00241-6
  [hep-th/0010113].

\bibitem{ChangYoung:2009sn}
  E.~Chang-Young, D.~Lee and Y.~Lee,
  ``Coordinate Dependence of Chern-Simons Theory on Noncommutative AdS(3),''
  arXiv:0812.3507 [hep-th].







\bibitem{Reisenberger:1996pu} 
  M.~P.~Reisenberger and C.~Rovelli,
  ``'Sum over surfaces' form of loop quantum gravity,''
  Phys.\ Rev.\ D {\bf 56}, 3490 (1997)
  doi:10.1103/PhysRevD.56.3490
  [gr-qc/9612035].
 
\bibitem{torsion_BH}
V.~de Sabbata, Dingxiong Wang, and C.~Sivaram, ``Torsion effects in black hole evaporation,''
Annalen der Physik 502.6 (1990): 508-510.

  
\bibitem{Ciric:2016isg} 
M.~Dimitrijević Ćirić, B.~Nikolić and V.~Radovanović,
``Noncommutative $SO(2,3)_\star$ gravity: Noncommutativity as a source of curvature and torsion,''
Phys.\ Rev.\ D {\bf 96}, no. 6, 064029 (2017)
doi:10.1103/PhysRevD.96.064029
[arXiv:1612.00768 [hep-th]].

\bibitem{Marculescu:2008gw} 
  S.~Marculescu and F.~Ruiz Ruiz,
  ``Seiberg-Witten maps for SO(1,3) gauge invariance and deformations of gravity,''
  Phys.\ Rev.\ D {\bf 79}, 025004 (2009)
  doi:10.1103/PhysRevD.79.025004
  [arXiv:0808.2066 [hep-th]].


\bibitem{Kibble:1961ba} 
  T.~W.~B.~Kibble,
  ``Lorentz invariance and the gravitational field,''
  J.\ Math.\ Phys.\  {\bf 2}, 212 (1961).
  doi:10.1063/1.1703702

\bibitem{Torsion_gravity}
  F.~W.~Hehl, P.~Von Der Heyde, G.~D.~Kerlick and J.~M.~Nester,
  ``General Relativity with Spin and Torsion: Foundations and Prospects,''
  Rev.\ Mod.\ Phys.\  {\bf 48} (1976) 393;\\
  V.~De Sabbata and M.~Gasperini,
  ``Introduction To Gravity,''
  Singapore, Singapore: World Scientific (1985) 346p

\end{thebibliography}
 \end{document}